\documentclass[preprint,aps,amsmath,superscriptaddress,nofootinbib]{revtex4}
\usepackage{bm}
\usepackage{epsfig}

\newcommand{\nn}{\nonumber} 

\newcommand{\bea}{\begin{eqnarray}}
\newcommand{\eea}{\end{eqnarray}}

\newcommand{\mpi}{m_{\pi}}

\newcommand{\mdds}{\mu_{DD*}}

\newcommand{\Pb}{\bar{P}}
\newcommand{\Vb}{\bar{V}}
\newcommand{\Lchi}{\Lambda_{\chi}}

\begin{document}

\preprint{INT-PUB-11-042}


\title{The decay of the X(3872) into $\chi_{cJ}$ and the Operator Product Expansion in XEFT} 

\author{Sean Fleming\footnote{Electronic address:     fleming@physics.arizona.edu}}
\affiliation{Department of Physics, 
        University of Arizona,
	Tucson, AZ 85721
	\vspace{0.2cm}}

\author{Thomas Mehen\footnote{Electronic address: mehen@phy.duke.edu}}
\affiliation{Department of Physics, 
	Duke University, Durham,  
	NC 27708\vspace{0.2cm}}

\date{\today\\ \vspace{1cm} }


\begin{abstract}
XEFT is a low energy effective theory for the X(3872) that can be used to systematically analyze the decay and production
of the X(3872) meson, assuming that it is a weakly bound state of charmed mesons. In a previous paper, we calculated 
the decays of  X(3872) into $\chi_{cJ}$ plus pions using a two-step procedure in which Heavy Hadron Chiral Perturbation 
Theory (HH$\chi$PT) amplitudes are matched onto XEFT operators and then X(3872) decay rates are then calculated using 
these operators. The procedure leads to IR divergences in the three-body decay $X(3872) \to \chi_{cJ} \pi \pi $ when virtual $D$ mesons can go on-shell in tree level HH$\chi$PT diagrams. In previous work, we regulated these IR divergences with the $D^{*0}$ width. In this work, 
we carefully analyze $X(3872) \to \chi_{cJ} \pi^0$ and $X(3872) \to \chi_{cJ} \pi \pi$ using 
the operator product expansion (OPE) in XEFT.  Forward scattering amplitudes in HH$\chi$PT are matched onto local
operators in XEFT, the imaginary parts of which are responsible for the decay of the $X(3872)$.  Here we show that the IR divergences are regulated by the binding momentum of the $X(3872)$ rather than the width of the $D^{*0}$ meson. In the OPE, these IR divergences cancel in the calculation of the matching coefficients so the correct predictions for the $X(3872) \to \chi_{c1} \pi \pi$ do not receive enhancements due to the width of the $D^{*0}$. 
We give updated predictions for the decay $X(3872) \to \chi_{c1} \pi \pi$  at leading order in XEFT.

\end{abstract}

\maketitle

\newpage

\section{Introduction}

The X(3872)~\cite{ Choi:2003ue,Acosta:2003zx,Abazov:2004kp}
 is most likely a molecular bound state of neutral $D^0$ and $D^{*0}$ mesons (and their antiparticles). 
Recently we developed an effective field theory for the X(3872) called XEFT which can be used to systematically analyze its properties 
under this assumption~\cite{Fleming:2007rp}. This theory contains nonrelativistic $\pi^0$,  $D^0$, and $D^{*0}$ mesons as its degrees of freedom.
It has been used to rederive effective range theory predictions for $X(3872)\to D^0 \bar{D}^{0} \pi^0$ and to calculate the leading corrections to this process~\cite{Fleming:2007rp}, to calculate elastic scattering of charm mesons off the $X(3872)$~\cite{Canham:2009zq},  the cross section for $\pi^+ X(3872) \to D^{*+}D^{*0}$ ~\cite{Braaten:2010mg} and to calculate decays of $X(3872)$ to quarkonia~\cite{Fleming:2008yn,Mehen:2011ds}. Decays to final states with $\chi_{cJ}$ are interesting because heavy quark symmetry predicts the relative rates for decays to different $\chi_{cJ}$ and therefore the relative rates can be used to test the molecular interpretation of the state~\cite{Dubynskiy:2007tj}.

The purpose of this paper is to carefully revisit the calculation of three-body decays, $X(3872) \to \chi_{cJ} \pi \pi$, first performed in Ref.~\cite{Fleming:2008yn}. 
In the case of decays to $\chi_{c1}$, Ref.~\cite{Fleming:2008yn} found double poles coming from static $D$ meson propagators that can go on-shell in certain regions of three-body phase space. These double poles  lead to divergent phase space integrals in the total decay rate.  Since the divergence is associated with an internal $D$ meson going on-shell we will
refer to this as an IR divergence. Ref.~\cite{Fleming:2008yn} regulated these double poles by performing 
a matching calculation with $D^{*0}$ having a complex nonrelativistic energy $\Delta + i \Gamma_{D^{*0}}/2$, where $\Delta$ is the hyperfine
splitting between the $D^0$ and $D^{*0}$ and $\Gamma_{D^{*0}}\approx 68\, \rm {keV}$ is the width of the $D^{*0}$.   
When the double pole is regulated this way the decay rate is enhanced by a factor $2 \pi E_\pi/\Gamma_{D^{*0}} \sim 10^4$ where 
$E_\pi$ is the typical energy of the pion in the decay. Therefore, predictions for decay rates for $X(3872)\to \chi_{c1}\pi^0 \pi^0$ in Ref.~\cite{Fleming:2008yn} were found to be quite large. 
In this paper,  we argue that the IR divergence is not regulated by the width of the $D^{*0}$ but by larger mass scales. 
We show how to determine how to properly handle these IR divergences in these calculations and give the correct prediction for the three-body decay
$X(3872) \to \chi_{c1} \pi \pi$. The issues addressed in this paper should be relevant to other decays of the $X(3872)$ as well as the decays of other states that could be molecular bound states of hadrons, e.g., the recently discovered $Z_b$ states \cite{Collaboration:2011gja}. (For an XEFT-like treatment of these states, see Ref.~\cite{Mehen:2011yh}). 

XEFT is supposed to be valid near (within a few MeV) of the $D^0\bar{D}^{*0}$ threshold and other nearby thresholds, such as the $D^+ D^{*-}$ threshold which is about 8 MeV away are not explicitly included in the calculations mentioned above. A recent calculation of the line shape in the vicinity of the $X(3872)$ in Ref.~\cite{Baru:2011rs} shows that this is a very good approximation. It would be interesting to revisit many XEFT calculations with charged mesons included explicitly in the theory but we will not do so in this paper.

In order to understand the nature of the IR divergences we consider an alternative approach to calculating the decay rate
$X(3872) \to \chi_{cJ} \pi \pi$. We begin by reviewing how the hadronic decays to quarkonia plus pions were calculated in 
Ref.~\cite{Fleming:2008yn}, using $X(3872) \to \chi_{cJ} \pi^0$ to illustrate the technique.  The underlying short-distance 
process is  $D^0 \bar{D}^{*0} \to \chi_{cJ} \pi^0$.\footnote{Here and below charge conjugated processes are implied.}
By short-distance we mean that the process occurs on length scales much shorter than the typical separation of the constituents of the 
$X(3872)$. This length scale is not well known since the binding energy of the $X(3872)$ is not accurately determined, but is expected to be large compared to typical hadronic scales because the $X(3872)$ is very weakly bound. The wave function of the 
$D^0$-$\bar{D}^{*0}$ at long-distances is 
\bea\label{wfn}
\psi_{D D^*}(r)  \sim \frac{e^{-\gamma r}}{r} \, ,
\eea
where $\gamma = \sqrt{2 \mu_{D D^*} B}$, where $\mu_{D D^*}$ is the reduced mass of the charmed mesons and $B$ 
is the binding energy. From the currently measured masses of the $X(3872), D^0$, and $D^{*0}$~\cite{Yao:2006px}
we know that $B = 0.42 \pm 0.39$ MeV. 
For this range of binding energies,  the root mean squared separation is $r_X = \hbar c/(\sqrt{2}\gamma) = 4.9^{+13.4}_{-1.4}$ fm. 
This is much larger than the size of the $D^0$ and $D^{*0}$ meson so the charm and anticharm quarks in the $X(3872)$ are well separated 
most of the time. To coalesce into a quarkonium the $D$ mesons must come together to within 1 fm, a scale much shorter than $\hbar c/\gamma$.
One then expects to be able to derive a factorization theorem~\cite{Braaten:2005jj, Braaten:2006sy,Braaten:2005ai}
which schematically takes the form
\bea\label{schematic}
\Gamma[X(3872)\to L.H.] \propto |\psi_{D D^*}(0)|^2 \sigma[D^0 \bar{D}^{*0} + c.c \to L.H.] \, ,
\eea
similar to what is found for the decay of quarkonia and other nonrelativistic bound states. However, the wavefunction in Eq.~(\ref{wfn}) diverges 
as $r\to0$ so Eq.~(\ref{schematic}) is not quite right. The description of the $X(3872)$ as a two-body bound state surely
fails as $r \to 0$. New channels such as $J/\psi \rho$ and $D^+ D^{*-} +c.c.$ open up and at some point a hadronic description
in terms of mesons ceases to work entirely. In XEFT this is dealt with by requiring that the cutoff, $\Lambda$, be taken 
not too much larger than a few times $\gamma$ so these effects are integrated out and included as short distance operators in XEFT.
Thus we expect $|\psi_{D D^*}(0)|^2$ to be replaced by the appropriate short-distance matrix element in XEFT. 

\begin{figure}[t]
 \begin{center}
 \includegraphics[width=2.5in]{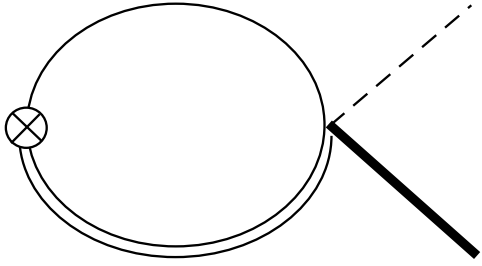}
 \end{center}
 \vskip -0.7cm \caption{Feynman diagram contributing to the 
$X(3872) \to \chi_{cJ} \,\pi^0$ decay amplitude. The circle with cross represents the interpolating 
field for the $X(3872)$.  }
\label{Xchipi}
 \end{figure}

The factorization theorem of Ref.~\cite{Fleming:2008yn} is derived as follows: the amplitude for the transition $D^0 \bar{D}^{*0} \to \chi_{cJ} \pi^0$ is computed in Heavy Hadron Chiral Perturbation Theory (HH$\chi$PT)~\cite{Wise:1992hn,Burdman:1992gh,Yan:1992gz}.  The amplitude contains contributions from contact interactions and tree level graphs with virtual $D$ mesons.
For the two-body decays, the $D$ mesons are off-shell by $E_\pi+ c \Delta$, where $c =0$ or $\pm 1$ depending on the channel.
Since $E_\pi$ ranges from $305$ to $432$ MeV in these decays and $\Delta = 142$ MeV, the virtuality of the $D$ meson 
is  large compared to the relevant scales in XEFT and the virtual $D$ mesons should  be integrated out. The amplitude for 
$D^0 \bar{D}^{*0} \to \chi_{c0} \pi^0$, for example,  is reproduced in XEFT by a local operator coupling the four fields:
\bea\label{Lagdecay}
{\cal L} = i \frac{C_{\chi, 0}(E_\pi)}{4\sqrt{m_\pi}} \, \frac{2}{\sqrt{3}} \,
(V^i \, \Pb +\Vb^i \, P) \, \frac{ \nabla_i \pi^0}{f_\pi} \, \chi_{c0}^\dagger \, ,
\eea
Here $V^i$ are the fields for $D^{*0}$ meson fields and $P$ are the fields for the $D$ mesons. The function $C_{\chi,0}(E_\pi)$ is calculated in Ref.~\cite{Fleming:2008yn} and contains the energy dependence of the virtual $D$ mesons. The decay of the $X(3872)$ is then calculated using the Lagrangian in Eq.~(\ref{Lagdecay}).
The relevant Feynman diagram, depicted in Fig.~\ref{Xchipi},
gives the result (for arbitrary $J$)
\bea\label{decayrates}
\Gamma[X(3872)\to \chi_{cJ}\pi^0] = \frac{ \gamma (\Lambda_{\rm PDS} -\gamma)^2}{2\pi}  \, 
\frac{m_{\chi_{cJ}}}{m_X}\, \frac{p_\pi^3}{72 \pi f_\pi^2}\, \alpha_J \, |C_{\chi,J}(E_\pi)|^2 \, ,
\eea
where $\alpha_0 =4$, $\alpha_1 = 3$, and $\alpha_2=5$, the functions $C_{\chi,J}(E_\pi)$ are given in Ref.~\cite{Fleming:2008yn}, and the PDS subtraction scheme is used  to regulate the 
divergent integral appearing in the calculation of the rate. The first term in Eq.~(\ref{decayrates}) is equal to the following 
matrix element in XEFT
\bea\label{ldme}
 \frac{ \gamma(\Lambda_{\rm PDS} -\gamma)^2}{2\pi} 
= \frac{1}{3}\sum_\lambda |\langle 0| \frac{1}{\sqrt{2}}{\epsilon}_i(\lambda) 
\,(V^i \, \Pb +\Vb^i \, P) |X(3872,\lambda)\rangle|^2 \, .
\eea 
Thus we find that the decay rate factorizes directly into a matrix element which plays the role of the wavefunction at the origin
squared times the remaining factors which are calculable in HH$\chi$PT. The matrix element is not calculable in XEFT, and the 
sensitivity to short distance physics is evident from the quadratic dependence on the cutoff. 

In next section of the paper, we briefly discuss power counting in HH$\chi$PT and XEFT. In the following section, we will show how to rederive the previous  results for two-body $X(3872)$ decays using the operator product expansion (OPE). In the next section of the paper, we will apply the OPE to the three-body decays of the $X(3872)$ where we reproduce our previous results for the decays $X(3872) \to \chi_{cJ} \pi \pi$ for $J= 0$ and $2$, and provide the correct treatment of IR divergences for the case  $J=1$.

\section{ Power-counting}

Before applying the OPE  in matching HH$\chi$PT onto XEFT, we explain the power counting in XEFT and HH$\chi$PT. In HH$\chi$PT  the expansion parameter is $Q/\Lchi$, where $\Lchi = 4 \pi f_\pi \sim 1 \, \textrm{GeV}$. Here $Q$ is either $m_\pi$ or $p$ or $E$, where $p$ ($E$) are either a loop or external momentum (energy) that are taken to be of order $m_\pi$. In XEFT the scales $m_\pi$, $\Delta$, and similar scales have been removed from the theory and reside in short-distance coefficients. As a result the short-distance coefficients scale with powers of $m_\pi/\Lchi$ (where we have chosen to replace $Q$ with $m_\pi$ for bookkeeping purposes). In Ref.~\cite{Fleming:2007rp}, XEFT is formulated as an expansion in $Q_X/\Lambda_X$ where $Q_X \sim p_\pi \sim p_D \sim p_{\bar{D}} \sim \gamma \sim \mu =\sqrt{\Delta^2-m_\pi^2} =  44$ MeV,  \footnote{The scale $\mu$ appears in potential pion exchange graphs in XEFT~\cite{Fleming:2007rp}.}  and the unknown scale, $\Lambda_X$, sets the range of the convergence of XEFT  and is expected to be of order $m_\pi$. The XEFT operators can have a complicated energy or momentum scaling since both $\Lchi$ and scales of order $m_\pi$ can appear in the denominator of XEFT operator coefficients. However, the scaling of XEFT operators becomes simple when all momentum scales $Q_X$ are considered as being $m_\pi v_\pi$, where $v_\pi \lesssim 1/3$, for the purposes of power counting. The pion kinetic energy, as well as energies in loop integrals, are $O(Q_X^2/m_\pi)$, and therefore  are counted as $m_\pi v_\pi^2$. In this formulation of the power counting,  powers of $m_\pi$ from momenta and energy in the numerator cancel powers of $m_\pi$ in the denominator so that a generic XEFT operator scales as powers of $(m_\pi/\Lchi)^m v^n_\pi$, where the integers $m$ and $n$ positive. In this formulation, calculations in XEFT are a simultaneous expansion in $m_\pi/\Lambda_\chi$ and $v_\pi$.
Examples will be given below.

\section{Two-body decays and the Operator Product Expansion}

Strictly speaking, matching amplitudes in HH$\chi$PT onto local operators in XEFT as in Eq.~(\ref{Lagdecay}) does not 
adhere to the EFT paradigm because in two-body decays the energies of the $\chi_{cJ}$ and $\pi^0$ in the final state are  too large for the power counting of XEFT. 
In the decay $X(3872) \to \chi_{cJ} \pi^0$, the energy of the pion is 433 MeV, 347 MeV, and 305 MeV, for $J=0,1$ and $2$, respectively.
The pion is relativistic and therefore is not properly treated in XEFT. Likewise the momentum of the $\chi_{cJ}$ exceeds the size mandated by XEFT power counting. These particles should really be integrated out of XEFT and their effects incorporated into local XEFT operators.  

\begin{figure}[t]
 \begin{center}
 \includegraphics[width=6in]{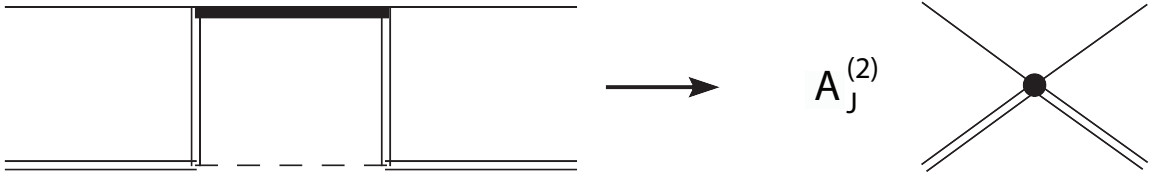}
 \end{center}
 \vskip -0.7cm \caption{The operator product expansion for $D^0 \bar{D}^{*0} \to \chi_{cJ} \pi^0 \to D^0 \bar{D}^{*0}$. The forward scattering matrix 
 element in HH$\chi$PT on the left-hand side  is matched onto products of short-distance coefficients, $A_J^{(2)}$, and XEFT operators on the right-hand side. }
\label{twobodyope}
 \end{figure}

The rigorous approach to $X(3872) \to \chi_{cJ} \pi^0$ is to use the OPE. In this approach the forward scattering matrix element for $D^0 \bar{D}^{*0} \to D^0 \bar{D}^{*0}$ via an intermediate $\chi_{cJ} \pi^0$ state is calculated in HH$\chi$PT, and matched onto products of short-distance coefficients and local operators in XEFT. This is illustrated in Fig.~\ref{twobodyope}. After matching, the decay rate for $X(3872) \to \chi_{cJ} \pi^0$ can be calculated in XEFT. This approach is conceptually similar to how short-distance annihilation decays of quarkonia  are treated in Non-Relativistic QCD (NRQCD)~\cite{Bodwin:1994jh}. 
In that theory, full QCD is used to calculate the matrix element squared for the processes $c\bar{c} \to gg,ggg$, for example. The matrix elements squared in full QCD determine the imaginary parts of coefficients of four-quark operators in NRQCD and the imaginary parts are responsible for the decay of the quarkonium state in NRQCD. 

\begin{figure}[t]
 \begin{center}
 \includegraphics[width=6in]{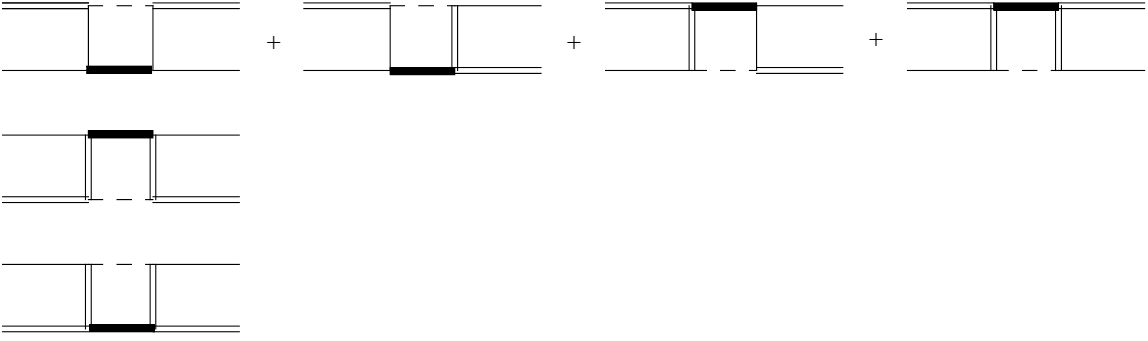}
 \end{center}
 \vskip -0.7cm \caption{Diagrams for the forward scattering amplitude for $D^0 \bar{D}^{*0} \to \chi_{cJ} \pi^0 \to D^0 \bar{D}^{*0}$ in HH$\chi$PT. The 
 top, middle, and bottom rows show the diagrams involving the $\chi_{c0}$,   $\chi_{c1}$, and $\chi_{c2}$, respectively.}
\label{twobodyfsme}
 \end{figure}

The HH$\chi$PT forward scattering matrix elements for $D^0 \bar{D}^{*0} \to D^0 \bar{D}^{*0}$ via intermediate $\chi_{cJ} \pi^0$ states 
are shown in Fig.~\ref{twobodyfsme} for the $\chi_{c0}$ (top row), $ \chi_{c1}$ (middle row), and $ \chi_{c2}$ (bottom row). The amplitudes for these diagrams are
\begin{eqnarray}
i{\cal M}^{(2)}_0 &= &\bigg( \frac{g g_1}{f_\pi} \bigg)^2 \int \frac{d^4 \ell}{(2 \pi)^4} \bigg( \frac{1}{\ell^0 -\Delta} +\frac{1}{3} \frac{1}{\ell^0 + \Delta}\bigg)^2 \frac{-|\vec{\ell}\, |^2}{\ell^2-m_\pi^2+i \epsilon} \frac{m_{\chi_0}}{m_X}\frac{1}{\Omega_0 -\ell^0+i\epsilon}\frac{m_{D^*} m_D}{2} \nn \\
i{\cal M}^{(2)}_1 &= & \frac{4}{3} \bigg( \frac{g g_1}{f_\pi} \bigg)^2 \int \frac{d^4 \ell}{(2 \pi)^4} \frac{-|\vec{\ell}\,|^2}{(\ell^0)^2}  \frac{1}{\ell^2-m_\pi^2+i \epsilon} 
\frac{m_{\chi_1}}{m_X}\frac{1}{\Omega_1 -\ell^0+i\epsilon}\frac{m_{D^*} m_D}{2} \nn \\
i{\cal M}^{(2)}_2 &= &  \frac{20}{9} \bigg( \frac{g g_1}{f_\pi} \bigg)^2 \int \frac{d^4 \ell}{(2 \pi)^4} \frac{-|\vec{\ell}\,|^2}{(\ell^0+\Delta)^2} \frac{1}{\ell^2-m_\pi^2+i \epsilon} 
\frac{m_{\chi_2}}{m_X}\frac{1}{\Omega_2 -\ell^0+i\epsilon}\frac{m_{D^*} m_D}{2} \,,
\label{hhchiptfsme}
\end{eqnarray}
where $\Omega_J = m_X-m_{\chi_{cJ}}$. The factor of $m_{\chi_{cJ}}/m_X$ is associated with each $\chi_{cJ}$ propagator.~\footnote{This factor can be understood by starting with a fully relativistic kinetic term for the $\chi_{cJ}$ fields,
\bea
\partial_\mu \chi_{cJ}^\dagger \partial^\mu \chi_{cJ} - m_{\chi_{cJ}}^2 \chi_{cJ}^\dagger  \chi_{cJ}  \, , \nn
\eea
then making the field redefinition
\bea
\chi_{cJ} \to \frac{1}{\sqrt{2 m_{\chi_{cJ}}}}    e^{- i m_X t} \chi_{cJ} \, . \nn
\eea
The factor of $\sqrt{2 m_{\chi_{cJ}}}$ ensures proper normalization of the non-relativistic $\chi_{cJ}$ field and the factor $m_X \approx m_{D^0} + m_{D^{*0}} $ in the exponent appears because all energies are measured relative to the $D^0 D^{*0}$ threshold. The factor $m_{\chi_{cJ}}/m_X$ could consistently be set to 1 within the accuracy of our calculation, but we keep it to obtain  expressions for decay rates that are identical to our earlier calculations.}
 To determine the two-body decay rate, we will need the discontinuity of the expressions above. The discontinuity can be found by cutting the both the pion and $\chi_{cJ}$ propagators, which amounts to the following replacement rules:
\begin{eqnarray}
 \frac{1}{\ell^2-m_\pi^2+i \epsilon} & \to &- 2 i \pi \, \delta(\ell^2-m_\pi^2)\\
 \frac{1}{\Omega_J -\ell^0+i\epsilon} & \to & -i \pi \, \delta(\Omega_J -\ell^0)\,.
 \end{eqnarray}
 Using these rules we find:
 \begin{eqnarray}
\frac{1}{2 i} \textrm{Disc}{\cal M}^{(2)}_0 &= & \frac{1}{8 \pi} \bigg( \frac{g g_1}{f_\pi} \bigg)^2 p_\pi^3  \bigg( \frac{1}{E_\pi -\Delta} +\frac{1}{3} \frac{1}{E_\pi + \Delta}\bigg)^2\frac{m_{\chi_0}}{m_X}\frac{m_{D^*} m_D}{2} \nn \\
\frac{1}{2 i} \textrm{Disc}{\cal M}^{(2)}_1 &= & \frac{1}{8 \pi} \frac{4}{3} \bigg( \frac{g g_1}{f_\pi} \bigg)^2  \frac{p_\pi^3}{E_\pi^2}\frac{m_{\chi_1}}{m_X}\frac{m_{D^*} m_D}{2}\nn \\
\frac{1}{2 i} \textrm{Disc}{\cal M}^{(2)}_2 &= & \frac{1}{8 \pi} \frac{20}{9} \bigg( \frac{g g_1}{f_\pi} \bigg)^2  \frac{ p_\pi^3}{(E_\pi+\Delta)^2}\frac{m_{\chi_2}}{m_X}\frac{m_{D^*} m_D}{2}\,,
\label{discfsme}
\end{eqnarray}
where in each expression $E_\pi = \Omega_J$ and $p_\pi = \sqrt{\Omega_J^2-m_\pi^2}$. 

Next we need the local XEFT operators which we will use in the matching: 
\begin{equation}\label{matcheq}
\delta {\cal L}^{(n)} = - \sum_J A^{(n)}_J (\bar{V}^i P + V^i \bar{P})^\dagger  (\bar{V}^i P + V^i \bar{P}) \,,
\end{equation}
where the subscsript $n$ refers to the number of particles in the final state and in this section we are focusing  on $n=2$.
The amplitudes for $D^0 \bar{D}^{*0} \to D^0 \bar{D}^{*0}$ for these operators are
\begin{eqnarray}
 -i  \sum_J A^{(2)}_J \, 4 m_{D^*} m_D \,,
\end{eqnarray}
and matching onto the HH$\chi$PT amplitudes in Eq.~(\ref{hhchiptfsme}) gives
\begin{eqnarray}
A^{(2)}_0 & = & -\frac{{\cal M}^{(2)}_0}{4 m_{D^*} m_D} \nn \\
A^{(2)}_1 & = & -\frac{{\cal M}^{(2)}_1}{4 m_{D^*} m_D} \nn \\
A^{(2)}_2 & = & -\frac{{\cal M}^{(2)}_2}{4 m_{D^*} m_D}  \,.
\label{twobodyampmatch}
\end{eqnarray}
The matching coefficients  scale as 
\begin{equation}
A^{(2)}_J  \sim g_1^2 \frac{\mpi}{\Lchi^2} \, ,
\end{equation}
in XEFT according to the power counting discussed earlier, since $E_\pi \sim p_\pi \sim \Delta \sim O(m_\pi)$.
Note that because the HH$\chi$PT forward scattering amplitudes match directly onto local four fermion operators in XEFT there is no suppression 
due to factors of $v_\pi$.

\begin{figure}[t]
 \begin{center}
 \includegraphics[width=3in]{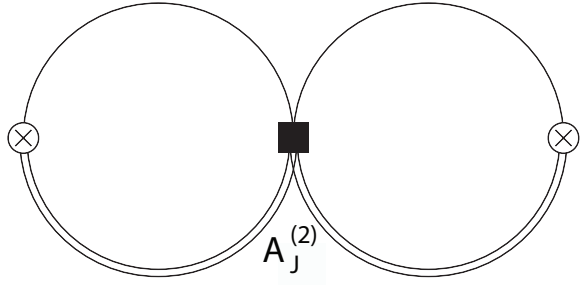}
 \end{center}
 \vskip -0.7cm \caption{The leading contribution to  $\Sigma_{A^{(2)}_J}(-E_X)$.}
\label{sigmaaj}
 \end{figure}

The XEFT decay rate for $X(3872)$ coming from final states with $n$ particles ($n-1$ pions) is given by 
\begin{equation}
\Gamma^{(n)}_J =  \frac{2 \, \textrm{Im} \,\Sigma_{A^{(n)}_J}(-E_X)}{ \textrm{Re}\, \Sigma'(-E_X)} \,,
\label{nbodydecayrate}
\end{equation}
where $ \Sigma(-E_X)$ is the sum of $C_0$ irreducible graphs contributing to the two-point function of the $X(3872)$ interpolating field, evaluated at the $X(3872)$ pole. 
$\Sigma_{A^{(n)}_J}(-E_X)$ is the contribution to $\Sigma(-E_X)$ from $C_0$ irreducible graphs with one insertion of  $A^{(n)}_J$. The leading order contribution to $\textrm{Re}\, \Sigma'(-E_X)$ was calculated in Ref.~\cite{Fleming:2007rp}:
\begin{equation}
\textrm{Re}\,\Sigma'(-E_X)_{LO} = \frac{\mdds^2}{2 \pi \gamma}\,.
\end{equation}
The leading contribution to $\Sigma_{A^{(2)}_J}(-E_X)$ is given by the diagram in Fig.~\ref{sigmaaj}, 
which evaluates to
\begin{equation}
\Sigma_{A^{(2)}_J}(-E_X)_{LO} =- \frac{A^{(2)}_J}{C_0^2} \,.
\end{equation}
Using these results in Eq.~(\ref{nbodydecayrate}), the leading order XEFT decay rate for $X(3872) \to \chi_{cJ} \pi^0$ is
\begin{equation}
\Gamma^{LO}[X(3872) \to \chi_{cJ} \pi^0] = -\frac{8 \pi \gamma}{\mdds^2 C^2_0} \textrm{Im} \,A^{(2)}_J \,,
\label{twobodydecayratelo}
\end{equation}
where the extra factor of two comes from including not only the $(V^i\bar{P})^\dagger V^i\bar{P}$ operator, but also the $(\bar{V}^i P)^\dagger \bar{V}^i P$, $(\bar{V}^i P)^\dagger V^i\bar{P}$, and $(V^i\bar{P})^\dagger \bar{V}^i P$ operators, and a factor of $1/2$ from the squaring the factor of $1/\sqrt{2}$ in the X(3872) wavefunction. This expression makes it clear that we are only interested in $\textrm{Im} \,A_J= \textrm{Disc} A_J/(2 i)$, which at leading order can be read off of Eqs.~(\ref{discfsme}) and (\ref{twobodyampmatch}). Inserting these into Eq.~(\ref{twobodydecayratelo}) gives 
\begin{eqnarray}
\Gamma^{LO}_{XEFT} [X(3872) \to \chi_{c0} \pi^0] & = & \frac{4 \pi \gamma}{\mdds^2 C^2_0} \frac{g^2 g_1^2}{32 \pi f^2_\pi} \frac{m_{\chi_0}}{m_X}p_\pi^3\bigg( \frac{1}{E_\pi -\Delta} +\frac{1}{3} \frac{1}{E_\pi + \Delta}\bigg)^2 \nn \\
\Gamma^{LO}_{XEFT} [X(3872) \to \chi_{c1} \pi^0] & = & \frac{4 \pi \gamma}{\mdds^2 C^2_0} \frac{g^2 g_1^2}{24 \pi f^2_\pi} \frac{m_{\chi_1}}{m_X}\frac{p_\pi^3}{E^2_\pi} \nn \\
\Gamma^{LO}_{XEFT} [X(3872) \to \chi_{c2} \pi^0] & = & \frac{4 \pi \gamma}{\mdds^2 C^2_0} \frac{5 g^2 g_1^2}{72 \pi f^2_\pi} \frac{m_{\chi_2}}{m_X}\frac{p_\pi^3}{(E_\pi+\Delta)^2} \,.
\end{eqnarray}
This agrees with the results of our previous paper~\cite{Fleming:2008yn}. 

\section{three--body decays and the Operator Product Expansion}

\begin{figure}[t]
 \begin{center}
 \includegraphics[width=5.5in]{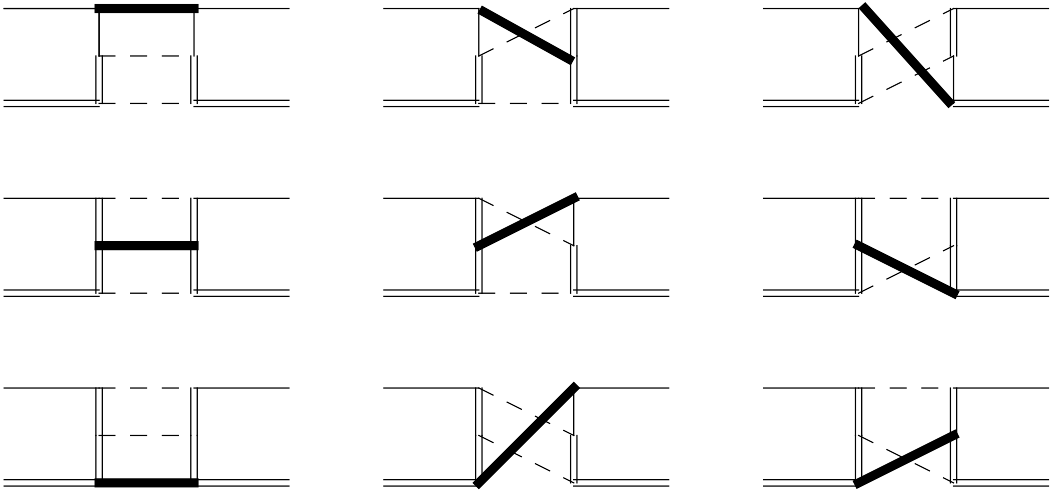}
 \end{center}
 \vskip -0.7cm \caption{HH$\chi$PT forward scattering matrix elements for $D^0 \bar{D}^{*0} \to D^0 \bar{D}^{*0}$ via intermediate $\chi_{cJ} \pi^0 \pi^0$ states, with $J=0,2$. The dark solid line is the  $\chi$ state.}
\label{threebodychi02fsme}
 \end{figure}

In this section we apply the OPE to the three-body decays of the $X(3872)$, which  is  straightforward for final states with a $\chi_{c0}$  or $\chi_{c2}$. The calculation of these decays is completely analogous to the two-body decays of the previous section. The decay rate for the three--body decays under consideration is given by Eq.~(\ref{nbodydecayrate}). The matching coefficient $A^{(3)}_J$ needed to calculate $\textrm{Im} \, \Sigma_{A^{(3)}_J}(-E_X)$
is determined  by matching HH$\chi$PT forward scattering matrix elements for $D^0 \bar{D}^{*0} \to D^0 \bar{D}^{*0}$ via intermediate $\chi_{cJ} \pi^0 \pi^0$ states
 onto XEFT . The diagrams involving $\chi_{c0}$  or $\chi_{c2}$ are shown in Fig.~\ref{threebodychi02fsme}. We calculate the imaginary parts of these graphs using the optical theorem to determine $\textrm{Im} \, A_J^{(3)}$.
The decay rate is then computed in terms of  $\textrm{Im} \, A_J^{(3)}$ by evaluating the diagram in Fig.~\ref{sigmaaj}. The final results for the LO decay rates for $X(3872) \to \chi_{cJ}\pi^0 \pi^0$, $J=0$ and $2$, are
\bea\label{xeftdr1}
\Gamma[X(3872) \to \chi_{c0} \pi^0 \pi^0] & = & 
\frac{2 \pi \gamma}{\mu_{DD^*}^2 C_0(\Lambda_{\rm PDS})^2}\frac{2\pi}{27}\frac{g^4 g_1^2 }{\Lambda_\chi^4}
\,   \frac{m_{\chi_{c0}}}{m_X} \\
& &  \hspace{-20 ex} \times \int^{\Omega_0 -\mpi}_{\mpi} d E_1 \int^{\Omega_0 -\mpi}_{\mpi} d E_2
\, \delta(E_1+E_2-\Omega_0) \, p_1^3 \, p_2^3 \,F^2(E_1,E_2)\,, \nn
\eea
where
\bea
F(E_1,E_2) = (E_2-E_1)\!\!\!\!
&&\left[\frac{1}{(E_1+\Delta)(E_2+\Delta)(E_1+E_2+\Delta)}  \right.\nn \\
&&\left.
+\frac{3}{E_1 \, E_2 \,(E_1+E_2-\Delta)} 
+\frac{\Delta}{E_1 \, E_2\, (E_1+\Delta)(E_2+\Delta)}\right] \, .
 \eea
and 
\bea\label{xeftdr3}
\Gamma[X(3872) \to \chi_{c2} \pi^0 \pi^0] & = & 
\frac{2 \pi \gamma}{\mu_{DD^*}^2 C_0(\Lambda_{\rm PDS})^2}\frac{40\pi}{27}\frac{g^4 g_1^2 }{\Lambda_\chi^4}
\,   \frac{m_{\chi_{c2}}}{m_X} \\
& & \hspace{-20 ex} \times \int^{\Omega_2 -\mpi}_{\mpi} d E_1 \int^{\Omega_2 -\mpi}_{\mpi} d E_2
\, \delta(E_1+E_2-\Omega_2) \, \nn \\
& &\hspace{-20 ex} \times  p_1^3 \, p_2^3
\bigg[C(E_1,E_2) ^2 + C(E_1,E_2)\big(D(E_1,E_2) -D(E_2,E_1)\big) \nn \\
& &\hspace{-10 ex}
+D(E_1,E_2)^2 + D(E_2,E_1)^2 + D(E_1,E_2) \,D(E_2,E_1)\bigg]
\,. \nn
\eea
where
\bea\label{dr6}
C(E_1,E_2) &=& \frac{E_2-E_1}{(E_1+\Delta)(E_2+\Delta)(E_1 + E_2+\Delta)} \, , \nn \\
D(E_1,E_2) &=& \frac{1}{(E_1+\Delta)E_2}  \, .
 \eea 
 Here, $E_i (\vec{p}_i)$ refers to the energy (three-momentum) of one of the $\pi^0$ and $p_i 
= |\vec{p}_i|$. The partial decay rates are symmetric under $1 \leftrightarrow 2$. In these decays
$\Delta$ is equal to the neutral hyperfine splitting, 
$\Delta= m_{D^{*0}}-m_{D^0} = 142 \, {\rm MeV}$. 
 These expressions for the decay rates are identical to those calculated in Ref.~\cite{Fleming:2008yn}, except for how the phase space integration is performed.  Ref.~\cite{Fleming:2008yn} calculated the rates with the exact relativistic phase space for the $\chi_{cJ}$ and pions. In the present paper we drop terms of order $m_\pi/M_X$ in the argument  of the energy-momentum conserving delta-function in three-body phase space.\footnote{Ref.~\cite{Fleming:2007rp} states that using the fully relativistic phase space is necessary because expanding in $m_\pi/M_X$ in the argument   of the energy/momentum conserving delta-function  would lead to unconstrained momentum integrals, but this is incorrect.} This considerably simplifies the calculation  of three-body phase space, which is now  a one-dimensional integral due to the delta-functions in Eqs.~(\ref{xeftdr1}) and (\ref{xeftdr3}). The rates calculated with this approximation for the phase space agree with the rates in our previous calculation to an accuracy of 10-15\%.

\begin{figure}[t]
 \begin{center}
 \includegraphics[width=5.5in]{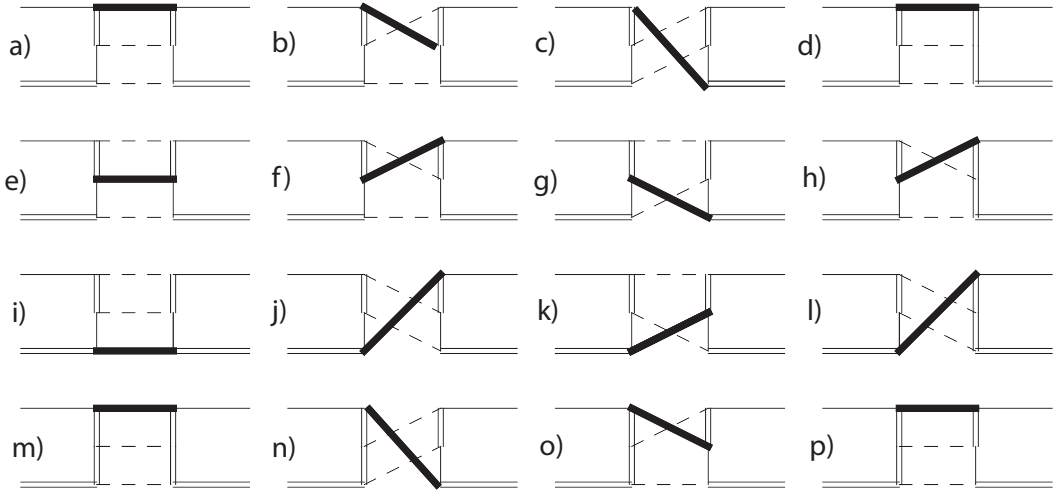}
 \end{center}
 \vskip -0.7cm \caption{HH$\chi$PT forward scattering matrix elements for $D^0 \bar{D}^{*0} \to D^0 \bar{D}^{*0}$ via intermediate $\chi_{c1} \pi^0 \pi^0$ states. The dark solid line is the  $\chi_{c1}$. }
\label{threebodychi1fsme}
 \end{figure}

Next we turn our attention to the problematic decay rate for $X(3872) \to \chi_{c 1} \pi^0 \pi^0$. The HH$\chi$PT diagrams for the forward scattering amplitude for $D^0 \bar{D}^{*0} \to D^0 \bar{D}^{*0}$ with an intermediate 
 $\chi_{c1}\pi^0 \pi^0$ state are shown in Fig.~\ref{threebodychi1fsme}.
 One would like to directly match these diagrams onto a local XEFT operator and obtain the coefficient ${\rm Im}\,A_1^{(3)}$ to calculate the decay rate as in our previous calculations. However, there are diagrams in which the internal $D$ meson propagators can go on-shell. Both internal $D$ meson propagators in Figs.~\ref{threebodychi1fsme}a), b), e), and f)  can go on-shell, leading to double poles in the three-body phase space integration. These double poles require a regulator to obtain finite answer for the total decay rate.  The internal $D$ meson propagators on the left-hand sides of Figs.~\ref{threebodychi1fsme}c), d), g), and h), and on the right-hand sides of Figs.~\ref{threebodychi1fsme}j), k), o), and p) can also go on-shell, which leads to single poles in the three-body phase space. These single poles can be dealt with a principal value prescription.  
 
 The imaginary part  of the diagrams in Fig.~\ref{threebodychi1fsme} is given by 
\begin{equation}
 \textrm{Im}\, {\cal M}^{(3)}_{1} = m_X \, \Gamma^{LO}_{\textrm{HH$\chi$PT} } [D^0 \bar{D}^{*0}\to \chi_{c1} \pi^0 \pi^0] \,.
 \label{imampdecraterel}
\end{equation}
where
\bea\label{chi1ddr}
& & \Gamma^{LO}_{\textrm{HH$\chi$PT} } [D^0 \bar{D}^{*0} \to \chi_{c1} \pi^0 \pi^0]  = 
\frac{\pi}{3}\frac{g^4 g_1^2  m_{\chi_{c1}} \mdds}{\Lambda_\chi^4}  i_1\\
& & i_1 = 
 \frac{2 }{m_X}\int^{\Omega_1 -\mpi}_{\mpi} d E_1 \int^{\Omega_0 -\mpi}_{\mpi} d E_2
\, \delta(E_1+E_2-\Omega_1) \,  p_1^3 \, p_2^3 \, \nn \\
& &\hspace{8 ex} \times 
\bigg\{ \frac{1}{3} \bigg[ 3 A(E_1,E_2)^2 - 2 A(E_1,E_2)\frac{1}{E_1 \, E_2}
+ 2\frac{1}{E_1 \, E_2}\frac{1}{(E_1 + E_2)^2}\bigg] + \frac{1}{E_1^2 \, E_2^2}\frac{E_1^2 +E_2^2}{(E_1 + E_2)^2}\nn\\
& &\hspace{12ex} 
+ \frac{2}{3}A(E_1,E_2) \bigg[ \frac{B_1(E_1,E_2)}{E_1 - \Delta}+ \frac{B_1(E_2,E_1)}{E_2 - \Delta}\bigg] \nn \\
& &\hspace{12ex} 
+ \frac{2}{3} \bigg[ \frac{B_1(E_1,E_2)}{E_1 - \Delta}  \frac{B_1(E_2,E_1)}{E_2 - \Delta}
-\frac{B_1(E_1,E_2)}{E_1 - \Delta} \frac{1}{E_1 }\frac{1}{E_1 + E_2}-\frac{B_1(E_2,E_1)}{E_2 - \Delta} \frac{1}{E_2 }\frac{1}{E_1 + E_2}\bigg] \nn \\
& &\hspace{12ex} 
-2 \frac{B_1(E_1,E_2)}{E_1 - \Delta} \frac{1}{E_2 }\frac{1}{E_1 + E_2} - 2 \frac{B_1(E_2,E_1)}{E_2 - \Delta} \frac{1}{E_1 }\frac{1}{E_1 + E_2}\nn \\
& &\hspace{12ex} 
+ \frac{B_1(E_1,E_2)^2}{(E_1 - \Delta)^2} + \frac{B_1(E_2,E_1)^2}{(E_2 - \Delta)^2} \bigg\} \, , \nn
\eea
where
\bea
A(E_1,E_2) &=& \frac{1}{E_1 \, E_2} +
\frac{E_1+E_2+2 \Delta}{(E_1 + E_2)(E_1+\Delta)(E_2+\Delta)} \, , \nn \\
B_1(E_1,E_2) &=& \frac{1}{E_2+\Delta} +
\frac{1}{E_1+E_2 }\, .
\eea
The terms in the curly brackets are increasingly singular when the $D$ meson goes on-shell, which occurs for either $E_1 \to \Delta$ or $E_2 \to \Delta$. Note that it is not kinematically possible to have both $E_1$ and $E_2$ go to $\Delta$ at the same time. In Ref.~\cite{Fleming:2008yn}, the integrals  were rendered finite by replacing $(E_i-\Delta)^2 \to (E_i-\Delta)^2+\Gamma^2/4$, where $\Gamma$ is the width of the $D^{*0}$. This yields a finite result that diverges in the $\Gamma \to 0$ limit. In what follows we will make a similar  replacement to render the integral in Eq.~(\ref{chi1ddr}) finite, but will interpret $\Gamma$ as a regulator rather than the physical width of the $D^{*0}$. 

\begin{figure}[t]
 \begin{center}
 \includegraphics[width=4in]{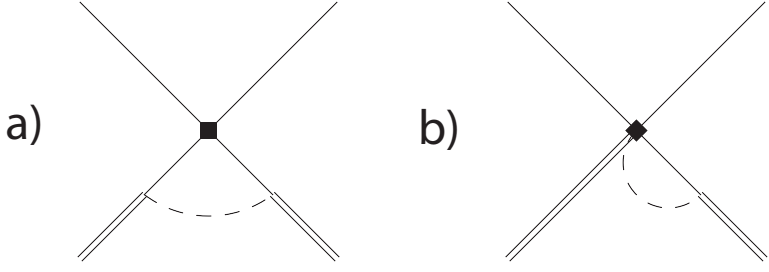}
 \end{center}
 \vskip -0.7cm \caption{Long distance XEFT contributions to the $X(3872) \to \chi_{c 1} \pi^0 \pi^0$ decay rate coming from diagrams with two on-shell $D$ meson propagators (on the left) or only one on-shell $D$ meson propagator (on the right).}
\label{xeftchiamps}
 \end{figure}
\begin{figure}[t]
 \begin{center}
 \includegraphics[width=6in]{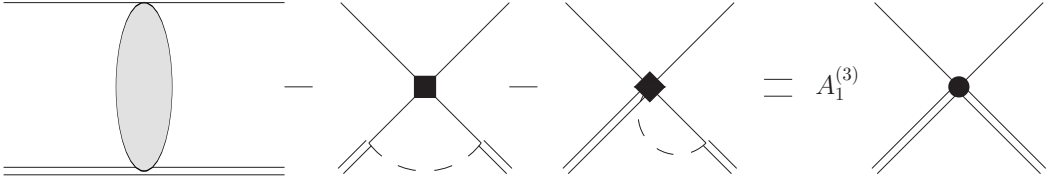}
 \end{center}
 \caption{The matching calculation for the coefficient $A_1^{(3)}$. The blob on the left represents the sum of all diagrams in Fig.~\ref{threebodychi1fsme}.}
\label{chi1match}
 \end{figure}

Because there are contributions to the graphs in Fig.~\ref{threebodychi1fsme} in which two $D$ mesons and the pion can go on-shell, these graphs cannot be directly matched onto a local XEFT operator. 
There are also long distance XEFT contributions which must be taken into account; these are shown in Fig.~\ref{xeftchiamps}.  
 Only after subtracting these long distance contributions from Eq.~(\ref{chi1ddr}) can the remainder term be matched onto the product of a local XEFT operator and the  short distance coefficient  $A^{(3)}_1$ in Eq.~(\ref{matcheq}). The diagrammatic equation for  the matching coefficient $A_1^{(3)}$ is shown in Fig.~\ref{chi1match}, where the blob on the left  represents all of the HH$\chi$PT forward scattering amplitudes in Fig.~\ref{threebodychi1fsme}.
The contact interaction represented by the solid square in Fig.~\ref{xeftchiamps}a) can be obtained by matching the HH$\chi$PT forward scattering matrix element for $D\bar{D} \to D\bar{D}$ via an intermediate $\chi_{c1} \pi$ state onto a local XEFT operator of the form
\begin{equation}
B (D \bar{D})^\dagger (D \bar{D}) \,.
\end{equation}
We will determine ${\rm Im}\, B$ below. There is also another contact interaction in Fig.~\ref{xeftchiamps}b) which is obtained by matching the HH$\chi$PT diagrams for $D^0\bar{D}^{0*} \to D^0 \bar{D}^0 \pi^0$.
We will not need the explicit form of this operator.

 According to the XEFT power counting discussed in Section II, the diagram in Fig.~\ref{xeftchiamps}a) scales as $(g^2_1 \mpi^3 v_\pi)/\Lchi^4$, which is $v_\pi \sim 1/3$ suppressed relative to  ${\rm Im}\, A_J^{(n)}$ (n=2,3).  This diagram has two $D$ meson propagators that can go on-shell, so it is IR divergent and therefore must be included in the matching calculation. We will see below that the IR divergence in Fig.~\ref{xeftchiamps}a) will  cancel the IR divergence in Eq.~(\ref{chi1ddr}),  rendering the coefficient $\textrm{Im} \, A_1^{(3)}$ finite. The power counting gives the correct estimate for the size of the remainder of the diagram, however, this remainder does not contribute to the $X(3872) \to \chi_{c1}\pi^0 \pi^0$ decay width.  The diagram in Fig.~\ref{xeftchiamps}b) only has a single on-shell  $D$ meson propagator and scales as  $(g^2_1 \mpi^3 v_\pi^3)/\Lchi^4$. Thus this contribution is suppressed by $v_\pi$ relative to the IR finite part of Fig.~\ref{xeftchiamps}a) and by $v_\pi^2$ relative to the coefficient $A_1^{(3)}$. Furthermore,  it is IR finite, i.e. it does not diverge as $\Gamma \to 0$,
 so we will drop this graph in the matching calculation as well as the calculation of the decay rate.

\begin{figure}[t]
 \begin{center}
 \includegraphics[width=5in]{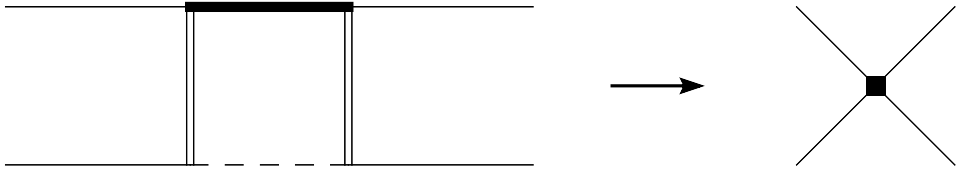}
 \end{center}
 \vskip -0.7cm \caption{Matching the HH$\chi$PT forward scattering matrix element for $D\bar{D} \to D\bar{D}$ via an intermediate $\chi_{c1} \pi$ state onto a local XEFT operator.}
\label{ddampmatch}
 \end{figure}

Next we will calculate the loop diagram in Fig.~\ref{xeftchiamps}a). We first calculate ${\rm Im}\,B$ which is determined by the matching diagrams shown in Fig.~\ref{ddampmatch}.
The HH$\chi$PT forward scattering matrix element for  $D\bar{D} \to D\bar{D}$ has the form
\begin{equation}
i {\cal M} = {\cal A}^{(D\bar{D} \to \chi_1 \pi)}_\textrm{HH$\chi$PT} m^2_D
\end{equation}
so 
\begin{equation}
B = -\frac{i}{4} {\cal A}^{(D\bar{D} \to \chi_1 \pi)}_\textrm{HH$\chi$PT} \,.
\end{equation}
The explicit expression for the HH$\chi$PT forward scattering matrix element calculated from the diagram on the left-hand side of Fig.~\ref{ddampmatch} is
\begin{equation}
i {\cal M} = m^2_D \frac{4 g^2 g^2_1}{f_\pi} \int \frac{d^4 \ell }{(2\pi)^4} \, \frac{|\vec{\ell}\,|^2}{(\ell^0 - \Delta)^2} \, \frac{1}{\ell^2 -\mpi^2 + i \epsilon} \, \frac{1}{\Omega_1-\Delta + \ell^0 + i \epsilon} \,,
\end{equation}
where we have summed over the four diagrams coming from assigning a $D$ and a $\bar{D}$ to each initial and final line. Once again we will only need the discontinuity of the above expression which results in a simple expression
\begin{equation}
\label{ddampdisc}
\textrm{Disc}\, {\cal A}^{(D\bar{D} \to \chi_1 \pi)}_\textrm{HH$\chi$PT}  =
- 4 \frac{g^2 g^2_1}{4 \pi f^2_\pi} \frac{m_{\chi_1}}{m_X} \frac{\big[(\Omega_1 - \Delta )^2 -\mpi^2 \big]^{3/2}}{\Omega^2_1}\,.
\end{equation}

Next we will evaluate the XEFT loop integral in the left diagram of Fig.~\ref{xeftchiamps}. The  amplitude for this diagram is 
\begin{equation}
i {\cal M} =  i B \frac{g^2}{12 f^2_\pi \mpi}  \int \frac{d^4 \ell}{(2\pi)^4} \, \frac{|\vec{\ell}\,|^2}{\ell^0 - \frac{\vec{\ell}^2}{2 \mpi} + \delta + i \epsilon}
\, \frac{1}{(\ell^0- i \epsilon)^2} \,,
\end{equation}
where $\delta = \Delta-\mpi$.
The energy integral can be carried out by method of contours, the angular integrals are trivial, and the resulting expression is
\bea\label{loopint}
i {\cal M} &=&  i B \frac{g^2 \mpi}{6 \pi^2 f^2_\pi}  \int_0^\Lambda d\ell \,\frac{\ell^{4}}{(\ell^2-\mu^2)^2+\hat{\Gamma}^2/4}  \nn \\
&=&  i B \frac{g^2 \mpi}{6 \pi^2 f^2_\pi}   \left( \Lambda +\pi \frac{ \mu^3}{\hat{\Gamma}} \right) \,,
\eea
where $\mu^2 = 2 \mpi \delta$, $\Lambda$ is a UV cutoff and we regulate the IR divergence by inserting a factor of $\hat{\Gamma}^2/4$, where $\hat\Gamma=\Gamma/(2m_\pi)$, into the $D$ meson propagator.
In this expression we have dropped all terms that vanish in the $\Lambda \to \infty$ limit We find the imaginary part of the amplitude by taking the discontinuity which, using the result of Eq.~(\ref{ddampdisc}),  gives
\bea
\label{singxeft}
\textrm{Im}\, {\cal M} & = &- i \,  \textrm{Disc} \, B \, \, \frac{g^2 \mpi}{12\pi^2 f^2_\pi}  \left( \Lambda +\pi \frac{ \mu^3}{\hat{\Gamma}} \right)\nn \\
& = &  \frac{16 \pi}{3}  \frac{g^4 g^2_1}{\Lchi^4} \frac{\big[(\Omega_1 - \Delta )^2 -\mpi^2 \big]^{3/2}}{\Omega^2_1} \frac{m_{\chi_1}}{m_X} \mpi
\left( \Lambda +\pi \frac{ \mu^3}{\hat{\Gamma}} \right)\,.
 \eea
To see the effect of subtracting the above result from the result in Eq.~(\ref{chi1ddr}), we will consider the $E_1 \to \Delta$ and $E_2\to \Delta$ limits of the most singular terms in 
Eq.~(\ref{chi1ddr}) (those on the last line of the equation), and compare it to Eq.~(\ref{singxeft}).  After converting from a decay rate to the imaginary part of the forward scattering amplitude, and accounting for the differences between the normalization of states in HH$\chi$PT and XEFT we find that the most singular terms in Eq.~(\ref{chi1ddr}) for energies near $\Delta$ have the form
\bea\label{div3bdy}
\textrm{Im}\, {\cal M} &\approx&  \frac{16 \pi}{3} \frac{g^4 g^2_1}{\Lchi^4} \frac{\big[(\Omega_1 - \Delta )^2 -\mpi^2 \big]^{3/2}}{\Omega^2_1} \frac{m_{\chi_1}}{m_X} \mpi
 \int_0^{\sqrt{\Omega_1(\Omega_1-2 \mpi)}} dp \,\frac{p^{4}}{(p^2-\mu^2)^2+\hat{\Gamma}^2/4}\,, \nn \\
 &=&   \frac{16 \pi}{3}  \frac{g^4 g^2_1}{\Lchi^4} \frac{\big[(\Omega_1 - \Delta )^2 -\mpi^2 \big]^{3/2}}{\Omega^2_1} \frac{m_{\chi_1}}{m_X} \mpi
\left( \pi \frac{ \mu^3}{\hat{\Gamma}} + ...\right) \, .
\eea
In the first line of Eq.~(\ref{div3bdy}) we have expressed the energy integral as a momentum integral to facilitate comparison with the loop integral in Eq.~(\ref{loopint}). The IR divergence from integrating over the double pole in three-body phase space is regulated by inserting the factor $\hat{\Gamma}^2/4$. This corresponds to exactly how the double poles were rendered finite in our calculation in Ref.~\cite{Fleming:2008yn}.
In the second line of Eq.~(\ref{div3bdy}) we keep only the IR divergence, the ellipsis represents contributions that are finite in the $\Gamma \to 0$ limit. It is clear that XEFT reproduces the IR divergence  in Eq.~(\ref{chi1ddr}). The $1/\Gamma$ enhanced contribution from the three-body phase space integral cancels out of the matching calculation, and does not contribute to the coefficient ${\rm Im} \,A_1^{(3)}$. Note there is a linear divergent term in the matching calculation that must removed by an XEFT counterterm.

Because matching calculations are insensitive to infrared physics they can usually be simplified by setting any physics on order of the infrared scale to zero. In the matching depicted in 
Fig.~\ref{chi1match} the IR energy scale is set by the typical XEFT energy scale: $E \sim \delta$. To make the IR terms  in Eq.~(\ref{chi1ddr}) obvious we shift the energies: $\tilde{E}_i = E_i - \mpi$. Then the singular terms become
\begin{equation}
\frac{1}{E_i -\Delta} \to \frac{1}{\tilde{E}_i -\delta}\,,
\end{equation}
and the limits of integration are ($\tilde{E}_i^\textrm{min} = 0, \tilde{E}_i^\textrm{max} = \Omega_1- 2 \mpi$). Setting the IR scale $\delta$ to zero further simplifies the singular terms:
\begin{equation}
\frac{1}{\tilde{E}_i -\delta} \to \frac{1}{\tilde{E}_i} \,.
\end{equation}
In addition, setting $\delta \to 0$ makes the XEFT loop integrals in Fig.~\ref{chi1match}  scaleless, and as a result they vanish. This greatly simplifies the matching calculation, which we do numerically. We find
\bea
\textrm{Im}\, A^{(3)}_1 & = & -\frac{\pi}{12} \frac{g^4g^2_1}{\Lchi^4} m_{\chi_1} i_1=  -\frac{\pi}{12} \frac{g^4g^2_1}{\Lchi^4} m_{\chi_1}\, (105 \, \textrm{MeV})^2 \, 
\eea
where $i_1$ is given in Eq.~(\ref{chi1ddr}).

The XEFT decay rate for $X(3872) \to \chi_{c 1} \pi^0 \pi^0$ can now  be calculated. The dominant contribution to the decay rate comes from the diagram in Fig.~\ref{sigmaaj} with an insertion of $A^{(3)}_1$.  The result is the same as that in of Ref.~\cite{Fleming:2008yn} but  with the $1/\Gamma$ enhanced terms omitted. In that paper we calculated the ratio of the branching fraction for the three-body decay  to the branching fraction for the LO two-body decay $X(3872) \to \chi_{c0} \pi^0$. We now find\footnote{A nearly identical result actually first  appeared in a preprint version of Ref.~\cite{Fleming:2008yn}. The calculations of that version of the paper are correct even though a proper  explanation of the prescription used to regulate the double poles is not given.}
\bea
\left( \frac{{\rm Br}[X(3872) \to \chi_{c1} \pi^0 \pi^0]}{{\rm Br}[X(3872) \to \chi_{c1} \pi^0]} \right)_{LO }= 2.9 \, 10^{-3}  \, .
\eea
Because of the small propagator denominators in the graphs for $X(3872) \to \chi_{c1}\pi^0 \pi^0$, this branching ratio is enhanced relative to the analogous branching ratio for $\chi_{c0}$ and $\chi_{c2}$ by 
$\sim (m_\pi/\delta)^2\sim 10^{2}$.

\begin{figure}[t]
 \begin{center}
 \includegraphics[width=3in]{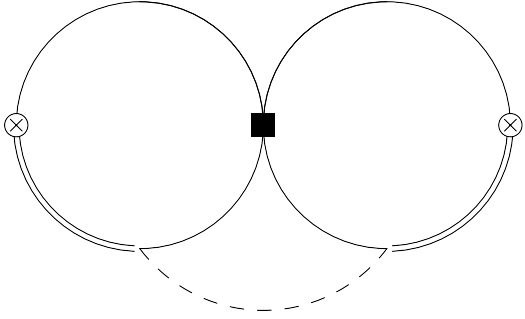}
 \end{center}
 \vskip -0.7cm \caption{A three loop XEFT diagram for $\Sigma(-E_X)$ with one insertion of $B$.}
\label{domxeftcont}
 \end{figure}

Fig.~\ref{domxeftcont} shows a three-loop diagram contributing to $\Sigma(-E_X)$   involving the $(D\bar{D})^2$ operator of Fig.~\ref{ddampmatch}.
After evaluating all three energy integrals by methods of contours, the expression for the amplitude in Fig.~\ref{domxeftcont} is
\bea
\label{oneloop}
i {\cal A}
&=&2 i  \, \frac{g^2 B}{12 f_\pi^2 \mpi} \int \frac{d^{3}k}{(2\pi)^{D-1}}  \, k^2\, \int \frac{d^{3}l}{(2\pi)^{3}} \, \frac{2\mdds}{l^2 +\gamma^2} \,\, \int \frac{d^{D-1}q}{(2\pi)^{D-1}} \frac{2\mdds}{q^{2} +\gamma^2} \\
& & \hspace{20 ex} \times 
\frac{m_D}{\frac{m_{D}}{2\mpi}(k^2-\mu^2)+\frac{1}{2} k^2- l\cdot k +l^2 +\frac{m_{D}}{2 \mdds}\gamma^2-i \epsilon}\nn \\
& & \hspace{20 ex} \times 
\frac{m_D}{\frac{m_{D}}{2\mpi}(k^2-\mu^2)+\frac{1}{2} k^2- q\cdot k +q^2 +\frac{m_{D}}{2 \mdds}\gamma^2-i \epsilon} \, , \nn
\eea
where $l$ is the loop three-momentum in the left-hand loop, $k$ is the loop three-momentum in the pion loop, and $q$ is the loop three-momentum in the right hand loop. In this expression $k\sim \mu \sim l \sim q \sim \gamma$, but $m_D \gg m_\pi$. This expression makes it clear that the double pole in XEFT is regulated by the binding momenta of the $X(3872)$, not the width of the $D^{*0}$, if  power suppressed terms are not dropped in the propagators appearing in the loop. But we can also drop terms suppressed by powers of $m_\pi/m_D$, and the double pole will still be regulated by the $i\epsilon$ prescription.
Dropping $m_\pi/ m_D$ suppressed terms in the propagator denominators, we find the amplitude factors into three integrals:
\bea
i {\cal A}
& \approx & i \, \frac{g^2 B}{12 f_\pi^2 \mpi}   \int \frac{d^{3}l}{(2\pi)^{D-1}} \, \frac{2\mdds}{l^2 +\gamma^2} \,\, \int \frac{d^{3}q}{(2\pi)^{3}} \frac{2\mdds}{q^{2} +\gamma^2} \nn \\ 
& & \times \int \frac{d^{3}k}{(2\pi)^{3}}  \, k^2\, \frac{(2\mpi)^2}{(k^2-\mu^2-i\epsilon)^2}\,.
\eea
The $l$ and $q$ integrals yield the factor $(\Lambda_{\rm PDS} -\gamma)/(2\pi)$ which, when combined with the wavefunction renormalization, yield the terms that are identified with the nonperturbative matrix element in Eq.~(\ref{ldme}). The $k$  integral is UV divergent and this divergence can be absorbed into an XEFT counterterm. The finite part of this integral is proportional to $\sqrt{-\mu^2 - i \epsilon}$, which is imaginary. 
If the coefficient $B$ is purely imaginary this cannot lead to a contribution to ${\rm Im}\, \Sigma(-E_X)$ and hence the width of $X(3872)$. There will be a contribution to ${\rm Im}\, \Sigma(-E_X)$ from the the real part of $B$. The real part of $B$ is due to elastic $D^0\bar{D}^0$ scattering so this contribution to ${\rm Im}\, \Sigma(-E_X)$ is not related to the decay $X(3872)\to \chi_{c1}\pi^0\pi^0$. It is rather a rescattering correction to the process  $X(3872)\to D^0 \bar{D}^0 \pi^0$.

\section{Conclusion}

In this paper we rigorously analyzed $X(3872)$ decays to $\chi_{cJ}$ plus pions using the OPE in XEFT. We reproduce our previous the results for decays to $\chi_{cJ} \pi^0$ for $J=0,1$, and 2 and $\chi_{cJ} \pi^0 \pi^0$ for $J=0$ and $2$.  Our previous calculations~\cite{Fleming:2008yn} for $\chi_{c1} \pi^0 \pi^0$ suffered from IR divergences due to double poles in the three-body phase space. These were previously  regulated with the width of the $D^{*0}$, $\Gamma$.  This led terms in the decay rate that are enhanced by $1/\Gamma$. In this paper, we argued that in XEFT these are IR divergences are regulated by the binding momentum of the $X(3872)$, not the width of the $D^{*0}$.  In the OPE analysis of the present paper, we find that the IR divergences cancel in the matching calculation and should not appear in the expression for the decay rate.

 \section{Acknowledgements}

We thank E. Braaten for helpful discussions. This work was supported in part by the  Director, Office of Science, Office of High Energy
Physics, of the U.S. Department of Energy under Contract No. DE-AC02-05CH11231U.S (SF), Office of Nuclear Physics, of the U.S. Department of Energy under grant numbers DE-FG02-06ER41449 (S.F.), 
DE-FG02-05ER41368 (T.M.), and DE-FG02-05ER41376 (T.M.).


\end{document}